\documentclass[a4paper]{article}

\usepackage{ISCSLP2024}
\usepackage{xcolor}
\usepackage{cite}
\usepackage{xcolor,hyperref,graphicx,subcaption}
\hypersetup{
    colorlinks=true,
    linkcolor=blue,
    filecolor=magenta,      
    urlcolor=red,
    citecolor=green
}
\title{Independent Feature Enhanced Crossmodal Fusion 
for Match-Mismatch Classification of Speech Stimulus and EEG Response \thanks{This work was partly supported by the project of the Ministry of Industry and Information Technology under Grant No.CBZ3N21-2}}

\name{Shitong Fan\textsuperscript{1,†}\thanks{† These authors contributed equally to this work.},
    Wenbo Wang\textsuperscript{2,†},
    Feiyang Xiao$^{1}$,
    Shiheng Zhang$^{1}$,
    Qiaoxi Zhu$^{3}$, and
    Jian Guan$^{1, *}$\thanks{* Corresponding author.}
    }
\address{
    $^1$Group of Intelligent Signal Processing (GISP), College of Computer Science and Technology,\\  Harbin Engineering University, Harbin, China\\
    $^2$Faculty of Computing, Harbin Institute of Technology, Harbin, China\\
    $^3$University of Technology Sydney, Ultimo, Australia}
\email{fanshitong@hrbeu.edu.cn, wwb1325864697@outlook.com, xiaofeiyang128@gmail.com,\\
        zhangshiheng@hrbeu.edu.cn, qiaoxi.zhu@gmail.com, j.guan@hrbeu.edu.cn} 
\begin{document}

\maketitle

\begin{abstract}
It is crucial for auditory attention decoding to classify matched and mismatched speech stimuli with corresponding EEG responses by exploring their relationship.
However, existing methods often adopt two independent networks to encode speech stimulus and EEG response, which neglect the relationship between these signals from the two modalities. 
In this paper, we propose an independent feature enhanced crossmodal fusion model (IFE-CF) for match-mismatch classification, which leverages the fusion feature of the speech stimulus and the EEG response to achieve auditory EEG decoding.
Specifically, our IFE-CF contains a crossmodal encoder to encode the speech stimulus and the EEG response with a two-branch structure connected via crossmodal attention mechanism in the encoding process, 
a multi-channel fusion module to fuse features of two modalities by aggregating the interaction feature obtained from the crossmodal encoder and the independent feature obtained from the speech stimulus and EEG response, 
and a predictor to give the matching result.    
In addition, the causal mask is introduced to consider the time delay of the speech-EEG pair in the crossmodal encoder, which further enhances the feature representation for match-mismatch classification.
Experiments demonstrate our method's effectiveness with better classification accuracy, as compared with the baseline of the Auditory EEG Decoding Challenge 2023.

\end{abstract}


\section{Introduction}

Electroencephalogram (EEG) is widely used to explore the mechanism by which the brain processes external information \cite{comprehensiveabiri2019, lstm_monesi, relatingpuffay2022, dreamdiffusionbai2023}. To analyze and interpret the brain's response to speech stimulus, auditory EEG decoding has recently attracted increasing attention due to its importance for the development of neuro-steered hearing aids \cite{attention_borsdorf2023}, such as the study on the match-mismatch classification to determine whether a given pair containing speech stimulus and EEG response correspond \cite{modeling}.

To address the match-mismatch problem, existing methods adopt various deep-learning-based structures to improve the representation of speech stimulus and EEG response for a better match and mismatch prediction. 
For example, the study in \cite{lstm_monesi} presents a model based on long short-term memory (LSTM)  \cite{lstmhochreiter1997}, which improves the accuracy of match-mismatch classification by considering variable delays in the EEG response to speech stimulus. 
The study in \cite{extractingmonesi2021} further explores the influence of using different levels of speech stimulus (i.e., envelope, voice activity, phoneme identity, and word embeddings) as the input for LSTM-based models in the match-mismatch task, and finds that mel-spectrogram can provide better classification performance.
Whereas the study \cite{enhancingsoman2023} introduces word boundary information to LSTM to account for the brain's discrete processing of speech stimulus, which further improves the classification performance.

Meanwhile, there are also some studies \cite{modeling, relatingpuffay2022, attention_borsdorf2023} that employ dilated convolutional neural networks \cite{Yu_Koltun_2015,wavenetoord2016} to capture the contextual information of both speech stimulus and EEG response,  expanding the receptive field while maintaining a low parameter count \cite{modeling}. 
Based on \cite{modeling}, the study in \cite{relatingpuffay2022}  introduces the fundamental frequency of speech and combines it with speech envelope, which improves the match-mismatch classification performance. 
Subsequently, \cite{attention_borsdorf2023} utilizes an attention encoder to replace the spatial convolution in \cite{modeling}, allowing for better exploration of the spatial and temporal feature of EEG response.
However, all these methods employ two separate networks to encode speech stimulus and EEG response, which neglects the relationship between these two modalities, leading to limited match-mismatch classification performance.

\begin{figure*}[!ht]
  \centering
  \includegraphics[scale=0.32]{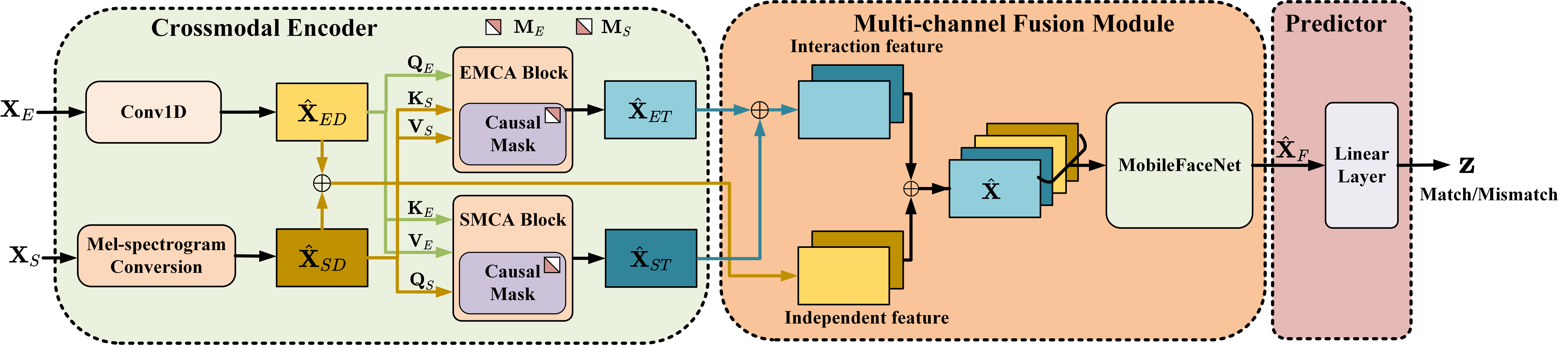}
  \caption{The model structure of our proposed Independent Feature Enhanced Crossmodal Fusion method (IFE-CF). ``$\bigoplus$" indicates the feature concatenate operation.  $\mathbf{X}_E \in \mathbb{R}^{D \times T}$ and $\mathbf{X}_S \in \mathbb{R}^{t}$ represent the EEG response and the speech stimulus, respectively. Here, $D$ denotes the number of EEG response channels, $T$ denotes the number of time frames in the EEG response, and $t$ represents the number of sampling points in the speech stimulus waveform. $\mathbf{M}_{E}$ and $\mathbf{M}_{S}$ are the causal mask matrices for EEG causal mask crossmodal attention (EMCA) block and speech causal mask crossmodal attention (SMCA) block,  respectively.}
  \label{fig:1}
\end{figure*}

In this paper, we propose an independent feature enhanced crossmodal fusion model (IFE-CF) for match-mismatch classification of auditory EEG decoding, which leverages the fusion feature of the speech stimulus and the EEG response to achieve speech-EEG pair classification.
Our IFE-CF consists of a crossmodal encoder, a multi-channel module and a predictor.  Specifically, we introduce a crossmodal attention mechanism \cite{multimodal} to build the crossmodal encoder, which can establish the connection between the latent features of speech stimulus and EEG response from two separate networks, to explore the relationship between the speech stimulus and EEG response. Furthermore, since EEG responses occur after a given moment of speech stimulus, to ensure that crossmodal attention focuses on the reasonable relationships when interacting with the latent features of both modalities, a causal mask strategy is introduced in crossmodal attention, which can filter out the unreasonable relationship in the interaction feature from the crossmodal encoder, thus improving the feature representation.

Then, a multi-channel fusion module is designed to fuse both the interaction feature from crossmodal attention encoder and the independent features from the separate networks. It provides a fine-grained feature that can reflect both the relationship between speech-EEG modalities and the specific characteristics of each individual modality for the predictor module, resulting in improved match-mismatch classification performance.  

Experiments are performed on the Auditory EEG Decoding Challenge 2023 dataset \cite{sparrkulee_accou2023}, and the results show that our proposed IFE-CF can outperform the official baseline method and surpass the Top-2 ranking system in Auditory EEG Decoding Challenge 2023. Ablation study and performance analysis are also conducted in our experiments, which validate the effectiveness of each component of our proposed method. 

The remainder of the paper is organized as follows: Section 2 presents our proposed method in detail; Section 3 shows the experimental results and analysis; and Section 4 concludes the paper and discusses potential future work.

\section{Proposed Method}

In this section, we present our proposed independent feature enhanced crossmodal fusion model (IFE-CF) in detail, which consists of a crossmodal encoder, a multi-channel fusion module, and a predictor for match-mismatch classification. The overall framework is given in Figure \ref{fig:1}.

\subsection{Crossmodal Encoder}

The crossmodal encoder adopts a two-branch structure to obtain the independent and interaction features of speech stimulus and EEG response via the independent and interaction feature encoding branches, respectively. 

\noindent\textbf{Independent Feature Encoding: }
For independent feature encoding, speech stimulus and EEG response are leveraged as the input of the crossmodal encoder module to obtain their independent features.

Inspired by work \cite{extractingmonesi2021}, a mel-spectrogram conversion is employed to convert the speech stimulus into the mel-spectrogram as the independent feature $\hat{\mathbf{X}}_{SD} \in \mathbb{R}^{d \times T}$ of speech stimulus, where $d$ and $T$ represent the dimension of extracted feature and the number of time frames, respectively. 

Meanwhile, a spatial convolution is utilized to obtain the independent feature of EEG response $\hat{\mathbf{X}}_{ED} \in \mathbb{R}^{d \times T}$, as follows
\begin{equation}
\hat{\mathbf{X}}_{ED} = \rm{Conv1D}(\mathbf{X}_{\mathit{E}}),
\end{equation}
where $\mathbf{X}_E \in \mathbb{R}^{D \times T}$ and $\rm{Conv1D}(\cdot)$ are the original input of EEG response and spatial convolution operation, respectively. $D$ denotes the channel number of the EEG response.

\noindent\textbf{Interaction Feature Encoding: }
The interaction feature encoding aims to explore the relationship between speech stimulus and EEG response according to their independent feature via two crossmodal attention blocks, i.e., EEG causal mask crossmodal attention (EMCA) block and speech causal mask crossmodal attention (SMCA) block. EEG causal mask crossmodal attention block is used to consider the EEG's relationship with speech in the encoding process of EEG response, while speech crossmodal attention block accounts for speech's relationship with EEG in the encoding process of speech stimulus.

Since these two crossmodal attention blocks perform in a similar way, we take the SMCA block as an example to present our method in detail for the sake of simplicity as follows.

In the speech causal mask crossmodal attention block, to obtain the interaction feature of speech stimulus $\hat{\mathbf{X}}_{ST} \in \mathbb{R}^{d \times T}$, the independent feature $\hat{\mathbf{X}}_{SD}$ and $\hat{\mathbf{X}}_{ED}$ is linearly transformed to obtain speech's query $\mathbf{Q}_{S}$, key $\mathbf{K}_{E}$, value $\mathbf{V}_{E}$,  as follows
\begin{equation}
    \left\{
\begin{aligned}
\mathbf{Q}_{S} &= \mathbf{W}_{Q, S} \hat{\mathbf{X}}_{SD}, \\
\mathbf{K}_{E} &= \mathbf{W}_{K, E} \hat{\mathbf{X}}_{ED}, \\
\mathbf{V}_{E} &= \mathbf{W}_{V, E} \hat{\mathbf{X}}_{ED},
\end{aligned}
\right.
\end{equation}
where $\mathbf{W}_{Q, S}, \mathbf{W}_{K, E}$, and $\mathbf{W}_{V, E} \in \mathbb{R}^{d_k \times d} $ are weight matrices, and $d_k$ is the dimension of these weight matrices.

Then, $\mathbf{Q}_{S}$, $\mathbf{K}_{E}$, and $\mathbf{V}_{E}$ are split into $h$ heads, and the $i$-th head's query $\mathbf{Q}_{s}^{i}$, key $\mathbf{K}_{e}^{i}$, and value $\mathbf{V}_{e}^{i}$ can be obtained, as follows
\begin{equation}
    \left\{
\begin{aligned}
\mathbf{Q}_{s}^{i} = \mathbf{W}_{_{Q, S}}^{i} \mathbf{Q}_{s}, \\
\mathbf{K}_{e}^{i} = \mathbf{W}_{K, E}^{i} \mathbf{K}_{e}, \\
\mathbf{V}_{e}^{i} = \mathbf{W}_{V, E}^{i} \mathbf{V}_{e},
\end{aligned}
\right.
\end{equation}
where $ \mathbf{W}_{Q, S}^{i}, \mathbf{W}_{K, E}^{i} $ and  $\mathbf{W}_{V, E}^{i} \in \mathbb{R}^{\frac{d_{k}}{h} \times d} $ are weight matrices. 

To consider the time delay between speech-EEG pair, a causal mask $\mathbf{M}_{S} \in \mathbb{R}^{T \times T} $ is introduced in the SMCA block to filter out the noise in the interaction feature calculation, which can be defined as
\begin{equation}
\mathbf{M}_{S} =
\begin{bmatrix}
0 & 0 & \cdots & 0 \\
 -\infty& 0 & \cdots & 0 \\
\vdots & \vdots & \ddots & \vdots \\
-\infty & -\infty & \cdots & 0\\
\end{bmatrix},
\end{equation}
Similar to the construction of the lower triangular mask matrix $\mathbf{M}_{S}$ in SMCA block, an upper triangular mask matrix $\mathbf{M}_{E} $ is also designed in EMCA block to filter out the irrelevant noise in the interaction feature.

Then, the interaction feature $\hat{\mathbf{X}}_{ST} $ can be obtained as follows
\begin{equation}
\hat{\mathbf{X}}^i_{ST} = \text{softmax}\left(\frac{\mathbf{Q}_{s}^{i} {\mathbf{K}_{e}^{i}}^\top}{\sqrt{d_k}} + \mathbf{M}_{S}\right)\mathbf{V}_{e}^{i},
\end{equation}
\begin{equation}
\hat{\mathbf{X}}_{ST} = \text{Concat}(\hat{\mathbf{X}}^1_{ST}, \ldots, \hat{\mathbf{X}}^h_{ST}) \mathbf{W}_O,
\end{equation}
where $\rm{softmax}(\cdot)$  and $\rm{Concat}(\cdot)$ are softmax function and concatenation operation respectively. 
$\mathbf{W}_O \in \mathbb{R}^{d \times d}$ is a learnable weight matrix. 

Finally, we can obtain four features including speech independent feature  $\hat{\mathbf{X}}_{SD}$, EEG independent feature  $\hat{\mathbf{X}}_{ED}$, speech interaction feature  $\hat{\mathbf{X}}_{ST}$ and EEG interaction feature  $\hat{\mathbf{X}}_{ET}$ via our two-branch crossmodal encoder.

\subsection{Multi-Channel Fusion Module}

The multi-channel fusion module is employed to further capture the relationship between speech stimulus and EEG response via deeply fusing the independent and interaction features of speech stimulus and EEG response.
Here, the independent feature of speech stimulus and EEG response are introduced in the fusion processing, which is used to ensure the fusion feature obtained by our multi-channel fusion module not only contains the relationship between speech stimulus and EEG response, but also includes the characteristics of these signals in their modality.

Thus, we concatenate $\hat{\mathbf{X}}_{SD}, \hat{\mathbf{X}}_{ED}, \hat{\mathbf{X}}_{ST}$ and $\hat{\mathbf{X}}_{ET}$ to obtain the fusion feature $\hat{\mathbf{X}} $ as follows
\begin{equation}
  \hat{\mathbf{X}} = \mathrm{Concat}(\hat{\mathbf{X}}_{SD}, \hat{\mathbf{X}}_{ED},  \hat{\mathbf{X}}_{ST}, \hat{\mathbf{X}}_{ET}),
  \label{eq2}
\end{equation} 
Then $\hat{\mathbf{X}}$ is fed into a MobileFaceNet structure \cite{MFN} to obtain the embedding feature $\hat{\mathbf{X}}_F$ for mach and mismatch prediction, as follows
\begin{equation}
  \hat{\mathbf{X}}_F = \text{MobileFaceNet}(\hat{\mathbf{X}}; \Theta),
  \label{eq2}
\end{equation} 
where $\Theta$ denotes the learnable parameters of MobileFaceNet. The detailed architecture of MobileFaceNet used in our method is given in Table \ref{tab:1}.

\subsection{Predictor}
To predict the match-mismatch result $\mathbf{z} \in \mathbb{R}^{1\times2}$, we apply a simple linear layer as our  predictor for match-mismatch prediction, expressed as follows
\begin{equation}
    \mathbf{z}= \text{softmax}(\mathbf{W} \hat{\mathbf{X}}_F + \mathbf{b}),
  \label{eqliner}
\end{equation}
where $\mathbf{W}$ and $\mathbf{b}$ denote the weight matrix and bias of the linear layer, respectively. 

\begin{table}[th]
  \caption{MobileFaceNet architecture used for feature embedding, where ef denotes the expansion factor, c is the number of output channels per layer, n represents the number of times each row operation is repeated, st is the stride.}
  \label{tab:1}
  \centering
  \resizebox{0.46\textwidth}{!}{
  \begin{tabular}{l@{}c c c c c c}
    \toprule
    \multicolumn{2}{c}{\textbf{Input}} & \textbf{Operator} & \textbf{ef} & \textbf{c} & \textbf{n} & \textbf{st} \\
    \midrule
    28$\times$192&$\times$\phantom{0}4 & conv3x3 & - & 64 & 1 & 2 \\
    14$\times$\phantom{0}96& $\times$\phantom{0}64 & depthwise conv3x3 & - & 64 & 1 & 1 \\
    14$\times$\phantom{0}96 & $\times$\phantom{0}64 & bottleneck & 2 & 64 & 5 & 2 \\
    \phantom{0}7$\times$\phantom{0}48& $\times$\phantom{0}64 & bottleneck & 4 & 128 & 1 & 2 \\
    \phantom{0}4$\times$\phantom{0}24& $\times$ 128 & bottleneck & 2 & 128 & 6 & 1 \\
    \phantom{0}4$\times$\phantom{0}24& $\times$ 128 & bottleneck & 4 & 128 & 1 & 2 \\
    \phantom{0}2$\times$\phantom{0}12& $\times$ 128 & bottleneck & 2 & 128 & 2 & 1 \\
    \phantom{0}2$\times$\phantom{0}12& $\times$ 128 & conv1x1 & - & 512 & 1 & 1 \\
    \phantom{0}2$\times$\phantom{0}12& $\times$ 512 & linear GDC conv7x7 & - & 512 & 1 & 1 \\
    \phantom{0}1$\times$\phantom{0}\phantom{0}1 & $\times$ 512 & linear conv1x1 & - & 128 & 1 & 1 \\
    \bottomrule
  \end{tabular}}
\end{table}

\section{Experiments and Results}
\subsection{Experimental Setup}

\noindent\textbf{Dataset:} 
We evaluate our method on the dataset of Auditory EEG Decoding Challenge 2023 \cite{sparrkulee_accou2023}, which collects EEG response data from 85 young participants with normal auditory systems. All these participants are native Dutch speakers. 
The speech stimuli are stories narrated by a single speaker whose native language is Flemish (Belgian Dutch). 
The training set contains the EEG responses of 71 participants, and the EEG responses of the remaining 14 participants are used as the evaluation set named \textbf{held-out subjects}. 
In addition, the EEG responses of these 71 participants, obtained by hearing new stories narrated by the speaker, are leveraged as another evaluation set named \textbf{held-out stories}.
In our experiments, the evaluation sets for \textbf{held-out subjects} and \textbf{held-out stories} are consistent with the evaluation sets used in Auditory EEG Decoding Challenge 2023. 

\noindent\textbf{Evaluation Metrics:}
To evaluate our method, the average accuracy \( Acc_{s} \) of the model's correct matched classification for subject \( s \) is calculated as follows
\begin{equation}
  Acc_{s} =  {\textstyle \sum_{i=0}^{n_{s}}[label_{predicted} = label_{true}]}/n_{s},
  \label{eq3}
\end{equation}
where \( n_{s} \) represents the number of samples. Then, the average accuracy of held-out subjects \( S_1 \), the average accuracy of held-out stories \( S_2 \) and the final overall score \( Score \) can be denoted as
\begin{equation}
  S_1 = \frac{1}{14} \sum_{s=72}^{85} ACC_s,
  \label{eq4}
\end{equation}
\begin{equation}
  S_2 = \frac{1}{71} \sum_{s=1}^{71} ACC_s,
  \label{eq5}
\end{equation}
\begin{equation}
  Score = \frac{2}{3} S_1 + \frac{1}{3} S_2,
  \label{eq6}
\end{equation}
where \( s \) denotes the subject identity number.

\noindent\textbf{Hyperparameters}:
All parameters in our model are optimized using the Adam optimizer \cite{Adam} with a learning rate of 1e-3 and a batch size of 64.

\begin{table}[th]
  \caption{Performance comparison with different methods, where held-out stories(\%), held-out subjects(\%), and score(\%) are used for evaluation.}
  \label{tab:2}
  \centering
  \resizebox{0.46\textwidth}{!}{
  \begin{tabular}{l c c c}
    \toprule
    \textbf{Methods} & \textbf{Held-out stories} & \textbf{Held-out subjects} & \textbf{Score} \\
    \midrule
    Baseline \cite{modeling} & 77.98	&78.55	&78.17 \\
    A-SM \cite{attention_borsdorf2023} & 79.98 & 78.17 & 79.38 \\
    \cmidrule(lr){1-4}
    Top-1 \cite{top1thornton2023relating}& \textbf{82.71} & \textbf{80.98} & \textbf{82.13} \\
    Top-2 \cite{top2borsdorf2023multi}& 79.61 & 77.93 & 79.05\\
    Top-3 \cite{top3cui2023relate}& 79.21 & 78.40 & 78.94\\
    \cmidrule(lr){1-4}
    IFE-CF (Ours) & \underline{80.82} & \underline{80.48} & \underline{80.71} \\
    \bottomrule
  \end{tabular}}
\end{table}

\subsection{Performance Comparison}

To show the effectiveness of our IFE-CF, we compare our IFE-CF with official baseline method (i.e., official Baseline \cite{modeling}), the top three ranking systems of Auditory EEG Decoding Challenge 2023  (i.e., Top-1 \cite{top1thornton2023relating}, Top-2 \cite{top2borsdorf2023multi} and Top-3 \cite{top3cui2023relate}), and a recent state-of-the-art method (i.e., A-SM \cite{attention_borsdorf2023}). The results are given in Table \ref{tab:2}. 

Here, \textit{held-out stories} represents the average accuracy of the model on the evaluation set for known subjects receiving unknown speech stimuli, and \textit{held-out subjects} denotes the average accuracy of the model on the evaluation set for unknown subjects. The \textit{Score} is the overall performance of the model on the evaluation set, which is the weighted sum of the average accuracy of the two parts. 

\begin{table}[th]
  \caption{Results of ablation study, where held-out stories, held-out subjects, and score are used for evaluation. }
  \label{tab:3}
  \small 
  \centering
  \resizebox{0.46\textwidth}{!}{
  \begin{tabular}{l c c c}
    \toprule
    \textbf{Methods} & \textbf{Held-out stories (\%)} & \textbf{Held-out subjects (\%)} & \textbf{Score (\%)} \\
    \midrule
    IFE-CF(w/o -D) & 78.67   &77.60	    &78.31\\
    IFE-CF(w/o -T) &79.60	 &80.32     &79.84 \\
    IFE-SF & 78.90 & 79.02 &78.94\\
    \cmidrule(lr){1-4}
    IFE-CF & \textbf{80.82} & \textbf{80.48} & \textbf{80.71} \\
    \bottomrule
  \end{tabular}}
   \vspace{-2mm}
\end{table}

Table \ref{tab:2} shows that our method outperforms the official baseline method and the recent state-of-the-art method while surpassing the Top-2 and Top-3 systems in Auditory EEG Decoding Challenge 2023, which is only slightly lower than the Top-1 system. Note that the Top-1 system is an ensemble system that integrates several systems with ensemble learning \cite{ensemblesagi2018}, whereas our proposed method is the only single system that achieved such performance for this task. The results demonstrate the effectiveness and superiority of our proposed method.

\subsection{Ablation Study}

To validate the effectiveness of the two-branch structure to exploit the interaction feature and independent feature from two modalities (i.e., speech stimulus and EEG response), we conduct an ablation study.
Here, IFE-CF(w/o -D) and IFE-CF(w/o -T)  represent our IFE-CF without independent feature and interaction feature in the fusing processing of multi-channel fusion module, respectively.
To better analyze the importance of the interaction and independent features while avoiding the influence of introducing a large number of learnable parameters in the attention calculation process, we replace the crossmodal attention mechanism with a self-attention mechanism in the EEG and speech causal mask crossmodal attention blocks, which is denoted as IFE-SF.  Results are given in Table \ref{tab:3}.

From Table \ref{tab:3}, we can see that the performance of our method IFE-CF without independent feature (i.e., IFE-CF(w/o -D) drops more significantly than that of  IFE-CF  without interaction feature (i.e., IFE-CF(w/o -T))  on all metrics (i.e., \textbf{held-out subjects} and \textbf{held-out stories}), which indicates that the independent feature contains more useful information that is not included in the interaction feature. Therefore, the classification performance can be enhanced by introducing the independent feature of speech stimulus and EEG response.

In addition, both IFE-CF without crossmodal attention mechanism (i.e., IFE-SF) and IFE-CF without interaction feature (i.e., IFE-CF(w/o -T)) do not consider the relationship between speech stimulus and EEG response in the encoding process. As a result, they perform worse than IFE-CF on evaluation metrics.
Moreover, the performance of IFE-SF is even worse than that of IFE-CF(w/o -T) in terms of all metrics. 
This indicates that without considering the relationship between speech stimulus and EEG response, 
simply introducing more learnable parameters or concatenating more independent features decreases the performance due to information redundancy.

Therefore, our IFE-CF can achieve better performance by fusing independent and interaction features in the multi-channel fusion module. This demonstrates the effectiveness of simultaneously exploring the relationship between these two modalities with interaction features and the specific characteristics of independent features from each modality.

\begin{figure}[t]
  \centering
  \includegraphics[scale=0.29]{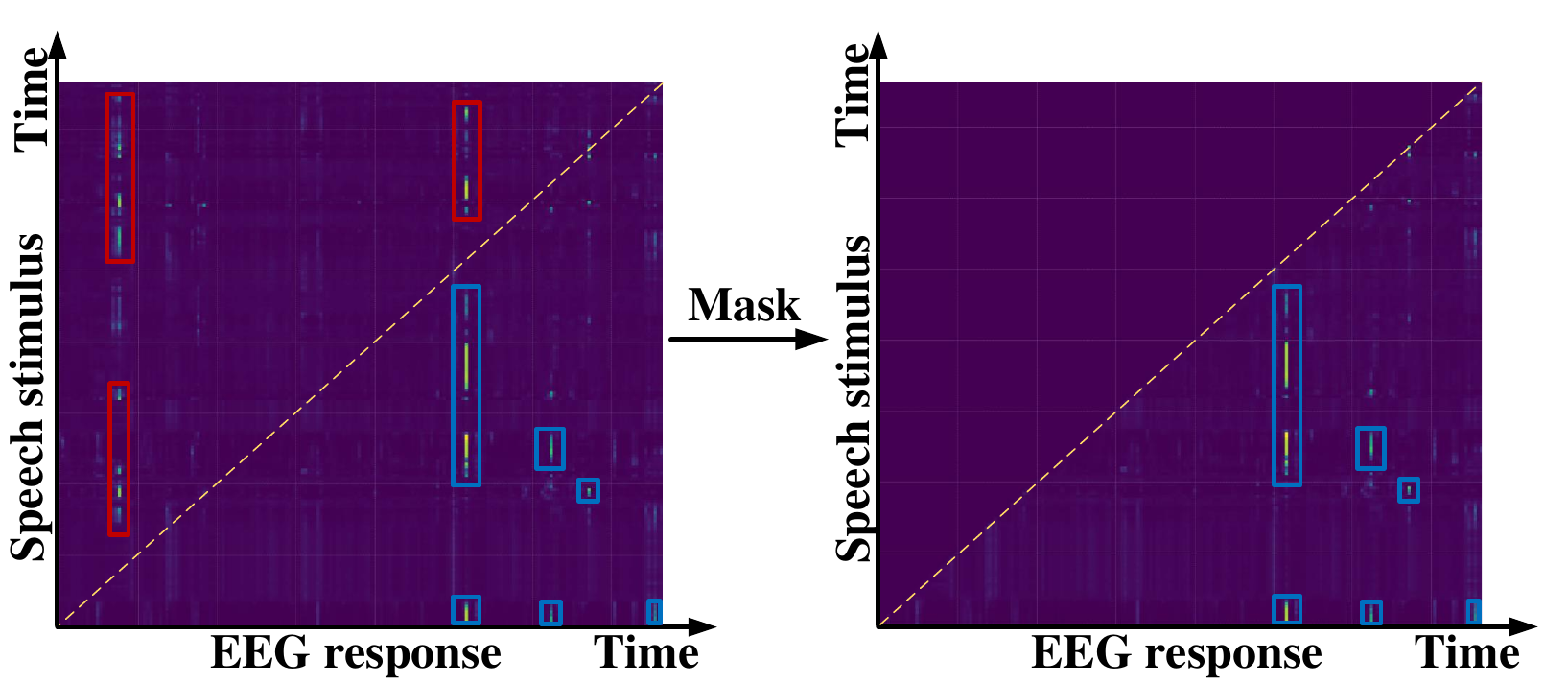}
  \caption{An example of applying causal mask in speech causal mask crossmodal attention (SMCA) block}
  \label{fig:2}
  \vspace{-5mm}
\end{figure}

\begin{table}[th]
  \caption{Ablation Study on Causal Mask, where held-out stories, held-out subjects, and score are used for evaluation.}
  \label{tab:5}
  \small 
  \centering
  \resizebox{0.46\textwidth}{!}{
  \begin{tabular}{l c c c}
    \toprule
    \textbf{Methods} & \textbf{Held-out stories (\%)} & \textbf{Held-out subjects (\%)} & \textbf{Score (\%)} \\
    \midrule
    IFE-CF(w/o -M) & 80.06&	\textbf{80.53}&	80.22 \\
    IFE-CF & \textbf{80.82} & 80.48 & \textbf{80.71} \\
    \bottomrule
  \end{tabular}}
   \vspace{-5mm}
\end{table}

\subsection{Analysis of Causal Mask}

We visually analyze the time delay characteristics of  matched speech-EEG pair by providing the heatmaps of the attention maps before and after applying  causal mask during crossmodal attention calculation. The results are given in Figure \ref{fig:2}.

Due to the time delay that exists between the speech-EEG pair, there is no dependency between the speech stimulus at a given moment and the EEG response preceding this moment. Therefore, by applying our causal mask strategy, the unreasonable parts (i.e., in red rectangles) are masked, while the reasonable parts that conform to the time delay (i.e., in blue rectangles) are retained, as shown in Figure \ref{fig:2}.

To further demonstrate the effectiveness of our causal mask strategy, we conduct the another ablative experiment. The results are given in Table \ref{tab:5}. Here, IFE-CF(w/o -M) denotes our IFE-CF without using causal mask strategy in the crossmodel blocks. We can see that the performance of IFE-CF(w/o -M) is worse than that of IFE-CF, which illustrates that highlighting the meaningful part of the crossmodal attention map by considering the time delay between EEG response and speech stimulus with causal mask strategy can further improve the match-mismatch classification performance.

\section{Conclusion}

In this paper, we have presented an independent feature enhanced crossmodal fusion model for match-mismatch classification. The crossmodal attention mechanism is utilized to build the connection between speech and EEG encoding process to capture the relationship between these two modalities. Furthermore, these modalities' independent and interaction features are fused in a multi-channel fusion module to explore the relationship between the two modalities and the specific characteristics of independent features from each modality, simultaneously. In addition, a causal mask strategy is introduced to account for the time delay of the matching speech-EEG pair, further improving the classification performance. Experimental results demonstrate the effectiveness and the superiority of our method.

\bibliographystyle{IEEEtran}

\bibliography{mybib}

\end{document}